\documentclass[twocolumn,preprintnumbers, amssymb,amsmath,aps,floatfix,prl,nofootinbib,superscriptaddress,showpacs]{revtex4-1}
\usepackage{epsfig}
\usepackage{bm}
\usepackage{amssymb}
\usepackage{amsmath}
\usepackage{color}
\usepackage{subfigure}
\usepackage[colorlinks,
            linkcolor=blue,
            anchorcolor=black,
            citecolor=blue
            ]{hyperref}

\newcommand{\bx}{\boldsymbol{x}}
\newcommand{\bb}{\boldsymbol{b}}
\newcommand{\br}{\boldsymbol{r}}
\newcommand{\bp}{\boldsymbol{p}}

\newcommand{\by}{\boldsymbol{y}}

\newcommand{\bk}{\boldsymbol{k}}

\newcommand{\be}{\begin{equation}}
\newcommand{\ee}{\end{equation}}
\newcommand{\bea}{\begin{eqnarray}}
\newcommand{\eea}{\end{eqnarray}}

\newcommand{\rTr}{\mathrm{Tr}}
\newcommand{\rT}{\mathrm{T}}
\newcommand{\rd}{\mathrm{d}}
\newcommand{\ri}{\mathrm{i}}

\allowdisplaybreaks[4]

\begin{document}
\title{On the elliptic flow of heavy quarkonia in $pA$ collisions}

\author{Cheng Zhang}\email{zhangcheng@mails.ccnu.edu.cn}
\affiliation{Key Laboratory of Quark and Lepton Physics (MOE) and Institute of Particle Physics, Central China Normal University, Wuhan 430079, China}

\author{Cyrille Marquet}\email{cyrille.marquet@polytechnique.edu}
\affiliation{CPHT, \'Ecole Polytechnique, CNRS, Universit\'e Paris-Saclay, F-91128 Palaiseau, France}

\author{Guang-You Qin}\email{guangyou.qin@mail.ccnu.edu.cn}
\affiliation{Key Laboratory of Quark and Lepton Physics (MOE) and Institute of Particle Physics, Central China Normal University, Wuhan 430079, China}

\author{Shu-Yi Wei}\email{shu-yi.wei@polytechnique.edu}
\affiliation{CPHT, \'Ecole Polytechnique, CNRS, Universit\'e Paris-Saclay, F-91128 Palaiseau, France}

\author{Bo-Wen Xiao}\email{bxiao@mail.ccnu.edu.cn}
\affiliation{Key Laboratory of Quark and Lepton Physics (MOE) and Institute
of Particle Physics, Central China Normal University, Wuhan 430079, China}
\affiliation{CPHT, \'Ecole Polytechnique, CNRS, Universit\'e Paris-Saclay, F-91128 Palaiseau, France}

\begin{abstract}
Using the dilute-dense factorization in the Color Glass Condensate framework, we investigate the azimuthal angular correlation between a heavy quarkonium and a charged light hadron in proton-nucleus collisions. We extract the second harmonic $v_2$, commonly known as the elliptic flow, with the light hadron as the reference. This particular azimuthal angular correlation between a heavy meson and a light hadron has first been measured at the LHC recently. The experimental results indicate that the elliptic flows for heavy-flavor mesons ($J/\psi$ and $D^0$) are almost as large as those for light hadrons. Our calculation demonstrates that this result can be naturally interpreted as an initial state effect due to the interaction between the incoming partons from the proton and the dense gluons inside the target nucleus. Since the heavy quarkonium $v_2$ exhibits a weak mass dependence according to our calculation, we predict that the heavy quarkonium $\Upsilon$ should have a similar elliptic flow as compared to that of the $J/\psi$, which can be tested in future measurements.
\end{abstract}
\maketitle

\textit{1. Introduction} Plenty of evidences for strong collectivity phenomenon in small collisional systems, such as $pp$ and $pPb$ collisions at the LHC and $dAu$ collisions at RHIC, have been reported\cite{Khachatryan:2010gv, CMS:2012qk, Abelev:2012ola, Aad:2012gla, Adare:2013piz, Adare:2014keg, Khachatryan:2015waa, Aidala:2018mcw} in the last few years. The collectivity in small systems is measured and computed in terms of particle azimuthal correlations in high multiplicity $pp$ and $pA$ collisions, and has become one of the most interesting and important topics in heavy ion physics. In these high multiplicity events, the azimuthal angular distributions of measured particle can be decomposed into Fourier harmonics with the corresponding coefficients $v_n \equiv \langle \cos n \Delta \phi \rangle$, where $\Delta \phi$ is the azimuthal angle difference between the measured particle and the reference particle or the reaction plane. 

In addition, recently there have been significant direct evidences that charm quarks also have sizable collectivity in small collisional systems. Both the ALICE\cite{Acharya:2017tfn} and CMS\cite{CMS:2018xac, Sirunyan:2018toe} collaborations have reported large values of elliptic flow $v_2$ for $J/\psi$ mesons and for $D^0$ mesons in $pPb$ collisions at the LHC, although they are slightly less than the $v_2$ values of light hadrons. 

One of the most successful explanations of the collectivity phenomenon in small collisional systems comes from the relativistic hydrodynamics approach. In this approach, the quark gluon plasma created in the collisions with high multiplicity are treated as relativistic fluids, and the flow harmonics can be viewed as the final-state effect due to hydrodynamic evolution of small collisional systems with certain amount of initial anisotropy. Excellent agreement\cite{arXiv:1304.3044,arXiv:1304.3403,1306.3439,arXiv:1307.4379,arXiv:1307.5060,arXiv:1312.4565,arXiv:1405.3605,Habich:2014jna, arXiv:1409.2160, arXiv:1609.02590,arXiv:1701.07145,arXiv:1801.00271} has been found between the hydrodynamics approach and the measured flow harmonics of light hadrons at both RHIC and the LHC. On the other hand, it is difficult for hydrodynamics to generate large collectivity for heavy mesons, since heavy quark in general does not flow as much as the light quark or gluon due to the large quark mass. Furthermore, recent calculation\cite{Du:2018wsj} also indicates that the observed $v_2$ from the ALICE and CMS collaborations can not come from final state interactions alone, since the final state interactions can only provide a small fraction of the observed $v_2$ for $J/\psi$ mesons. In addition, besides the hydrodynamics approach, there could be other possible mechanisms as suggested in Refs.\cite{Lin:2003jy,arXiv:1803.02072,arXiv:1804.02681, arXiv:1805.04081}.

Meanwhile, the Color Glass Condensate (CGC) framework or the saturation formalism shows that correlations between partons originated from the projectile proton and dense gluons inside the target nucleus, which can be written in terms of Wilson lines, can also provide significant amount of collectivity\cite{Armesto:2006bv, Dumitru:2008wn, Gavin:2008ev, Dumitru:2010mv, Dumitru:2010iy, Kovner:2010xk, Kovchegov:2012nd, Dusling:2012iga, Dumitru:2014dra, Dumitru:2014yza, Dumitru:2014vka, Lappi:2015vha, Schenke:2015aqa, Lappi:2015vta,McLerran:2016snu, Kovner:2016jfp, Iancu:2017fzn, Dusling:2017dqg, Dusling:2017aot, Fukushima:2017mko, Kovchegov:2018jun, Boer:2018vdi, Mace:2018vwq, Mace:2018yvl,Kovchegov:2013ewa,Altinoluk:2018ogz,Kovner:2018fxj,Kovner:2017ssr,Kovner:2018vec, Davy:2018hsl} for light hadrons. Usually this is regarded as the initial state effect prior to the onset of hydrodynamic evolution. The CGC framework has been quite useful in understanding the heavy quarkonium productions\cite{Ma:2014mri, Ma:2015sia, Watanabe:2015yca} in $pp$ and $pPb$ collisions in the low transverse momentum region. However, calculations on the $J/\psi$ $v_2$ in CGC framework are still lacking.

The objective of this paper is to study the elliptic flow harmonic $v_2$ of $J/\Psi$ mesons within a simplified model based on the Color Evaporation Model (CEM) and the dilute-dense factorization\cite{Dumitru:2002qt, Chirilli:2011km} in the CGC framework, and to demonstrate that a significant amount of $v_2$ can be generated due to the non-trivial QCD dynamics of the interaction between the partons from the proton projectile and dense gluons in the nuclear target. This calculation is a further extension of the two particle azimuthal correlation CGC calculation developed in Refs.~\cite{Lappi:2015vha, Lappi:2015vta, Dusling:2017dqg, Dusling:2017aot, Davy:2018hsl}. Besides, we need to consider the splitting of $c\bar c$ pair from a gluon ($g \to c\bar c$) in order to produce a $J/\psi$ meson in the final state. Similar to the measurements carried out at the LHC, we compute $v_2 \equiv \langle \cos 2 \Delta \phi \rangle$ based on the production of a $J/\psi$ meson in the CEM accompanied by another reference quark which eventually fragments into a charged hadron; i.e. we disregard the $q\to qg \to q J/\psi$ jet-like contributions.

The paper is organized as follows. In Sec.~2, we briefly introduce the framework employed in our calculation of $v_2$ for heavy quarkonia including the CGC correlators and the dilute-dense formalism for particle production as well as the CEM. In Sec.~3, we show the comparison between our numerical results and the LHC data, with some further comments. As a conclusion, the phenomenological implications of our model calculation are discussed in Sec.~4.

\textit{2. The elliptic flow of heavy quarkonia in $pA$ collisions} Let us now briefly mention the essential ingredients of the calculation which lead to the elliptic flow of heavy quarkonia in $pA$ collisions with a charged light hadron as the reference particle. In correspondence to high multiplicity events in $pA$ collisions, we assume that there are multiple active partons from the proton projectile participating the interaction with the target nucleus. To measure the $J/\Psi$ elliptic flow, a charged hadron is used as a reference particle in the LHC experiment. Similarly, to simplify the calculation, we pick a gluon and a quark from the proton with the quark serving as the reference while the gluon splitting into a pair of heavy quarks $\mathcal{Q}\mathcal{\bar Q}$, and compute their interactions with the target nucleus. We take into account all the possible correlations between the gluon (or $\mathcal{Q}\mathcal{\bar Q}$) and the reference quark generated by those interactions, up to $\frac{1}{N_c^2}$ order and neglect higher order corrections. As we show below, non-trivial color correlation starts to appear at the $\frac{1}{N_c^2}$ order which generates sizable elliptic flow for heavy quarkonia. We have numerically tested that the contributions from the $\frac{1}{N_c^4}$ order and other higher order terms are negligible in the region of interest where the transverse momentum is small. Our model is akin to the CGC model calculations\cite{Lappi:2015vha, Lappi:2015vta, Dusling:2017dqg, Dusling:2017aot, Davy:2018hsl} with the additional $g \to \mathcal{Q}\mathcal{\bar Q}$ splitting in order to produce heavy quarkonia in the final state.

\begin{figure}[tbp]
\begin{center}
\includegraphics[width=8.0cm]{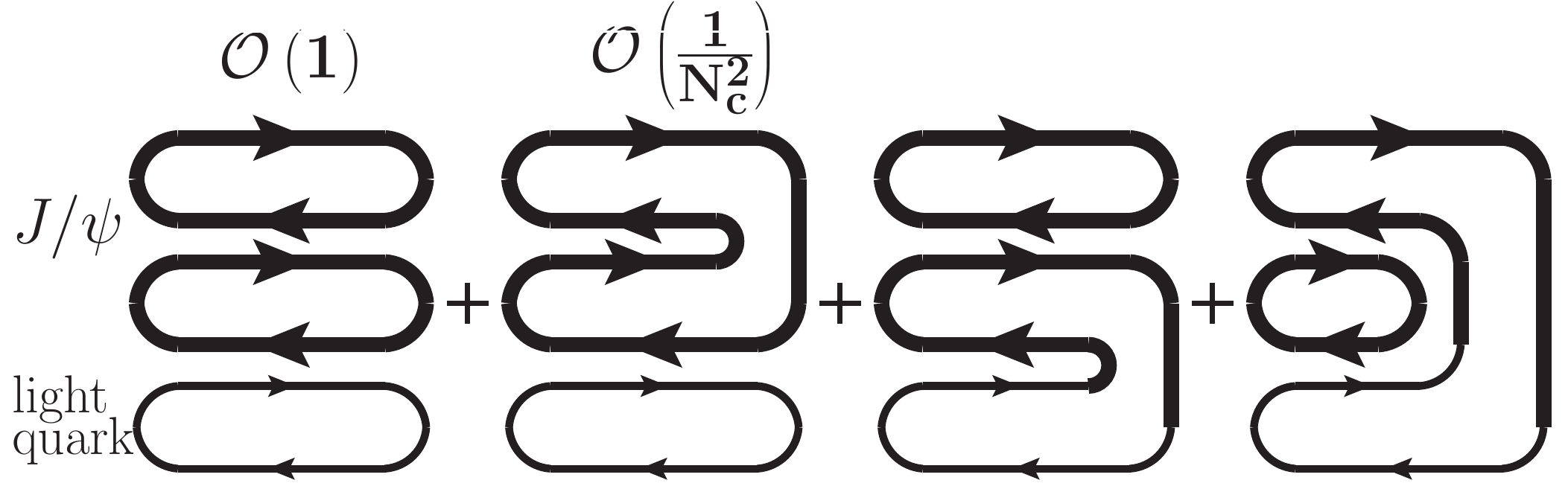}
\end{center}
\caption[*]{Illustration of the expectation value of three dipole correlatiors in the gluon background fields of the target nucleus. These four diagrams also show the origins of the each terms in Eq.~(\ref{ddd}). It is clear that only the last two diagrams contain azimuthal angular correlations between the produced $J/\psi$ and the reference light quark.}
\label{color-corr}
\end{figure}

Accompanied by a reference quark, the incoming gluon splits into a pair of heavy quarks $\mathcal{Q}\mathcal{\bar Q}$ before or after they traverse the dense nuclear target. Therefore, the differential spectrum of the production of $\mathcal{Q}\mathcal{\bar Q}$ and another light quark in the large $N_c$ limit can be written as\cite{Dominguez:2011wm}
\begin{widetext}
\begin{eqnarray}
\frac{\rd N^{gqA\rightarrow \mathcal{Q}\mathcal{\bar Q} qX}}{\rd^2\bk_1\rd^2\Delta\bk_1\rd^2\bk_2}&=&\mathcal{N}\int\frac{\rd^2\br\rd^2\br'}{(2\pi)^{2}}e^{-\ri\Delta\bk_1\cdot(\br-\br')}\prod_{i=1}^2\int\frac{\rd^2\bb_i\rd^2\br_i}{(2\pi)^{2}}
e^{-\ri\bk_i\cdot\br_i}\left<DDD\right>\psi(\br)\psi^*(\br'), \,
\end{eqnarray}
with the normalization factor $\mathcal{N}$ which cancels out when we compute $v_2$, and 
\begin{eqnarray}
DDD
\equiv [D(\bx_\mathcal{Q},\bx'_\mathcal{Q})D(\bx'_{\mathcal{\bar Q}},\bx_{\mathcal{\bar Q}})+D(\bx_g,\bx'_g)D(\bx'_g,\bx_g)-D(\bx_\mathcal{Q},\bx'_g)D(\bx'_g,\bx_{\mathcal{\bar Q}})-D(\bx'_{\mathcal{\bar Q}},\bx_g)D(\bx_g,\bx'_\mathcal{Q})]D(\bx_q,\bx'_q), \,\,\,
\end{eqnarray}
\end{widetext}
where the dipole correlators $D(\bx,\by)\equiv\frac{1}{N_c}\rTr U(\bx)U(\by)^\dag$. We denote $\bk_1$ as the transverse momentum of the $\mathcal{Q}\mathcal{\bar Q}$ pair and $\Delta\bk_1$ as the relative transverse momentum of $\mathcal{Q}\mathcal{\bar Q}$. $\bk_2$ stands for the transverse momentum of the reference light quark. In the above expression, $D(\bx_q,\bx'_q)$ corresponds to the reference quark production, while the other two color dipoles come from the heavy quark pair or the incoming gluon in the large $N_c$ limit. The full expression of the Wilson correlators for the $\mathcal{Q}\mathcal{\bar Q}$ production can be found in Ref.~\cite{Dominguez:2011wm} without taking the large $N_c$ limit. It is easy to see that the terms that we neglected above do not provide any correlation between the $\mathcal{Q}\mathcal{\bar Q}$ pair and the light reference quark. The transverse coordinates $\bx_{\mathcal{Q},\mathcal{\bar Q}, g, q}$ ($\mathcal{Q}, \mathcal{\bar Q}, g$ and the reference quark) inside the above dipole correlators can also be written as $\bx_\mathcal{Q}(\bx_{\mathcal{\bar Q}})\equiv\bx_g\pm\frac{\br}{2},\ \bx'_\mathcal{Q}(\bx'_{\mathcal{\bar Q}})\equiv\bx'_g\pm\frac{\br'}{2},\ \bx_g(\bx'_g)\equiv\bb_1\pm\frac{\br_1}{2},\ \bx_q(\bx'_q)\equiv\bb_2\pm\frac{\br_2}{2}$. Here the average $\langle DDD\rangle$ indicates the color averaging of three color dipoles in terms of the corresponding fundamental Wilson lines in the gluon background fields of target nuclei. The above four target averages can be computed in the McLerran-Venugopalan model\cite{McLerran:1993ni,McLerran:1993ka}, and non-trivial color correlations can appear when two color singlet dipoles are disconnected in order to form a singlet quadrupole due to inelastic exchanges with the target gluon fields as shown in Fig.~\ref{color-corr}. In the large $N_c$ limit, a general 3-dipole correlator can be cast into the following form up to the $\frac{1}{N_c^2}$ order\cite{Davy:2018hsl}

 \begin{widetext}
\begin{eqnarray}
\left<D(\bx_1,\bx'_1)D(\bx_2,\bx'_2)D(\bx_3,\bx'_3)\right>&=&e^{-\frac{Q_s^2}{4}[(\bx_1-\bx'_1)^2+(\bx_2-\bx'_2)^2+(\bx_3-\bx'_3)^2]} \nonumber
\\
&& 
\times \bigl[1+ F(\bx_1,\bx_1'; \bx_2,\bx_2' )+ F(\bx_3,\bx_3'; \bx_2,\bx_2' ) + F(\bx_1,\bx_1'; \bx_3,\bx_3' ) \bigr] , \label{ddd} \\
 \textrm{with} \quad F(\bx_1,\bx_1'; \bx_2,\bx_2')& =& \frac{[Q_s^2(\bx_1-\bx'_1)\cdot(\bx_2-\bx'_2)]^2}{4N_c^2} \int_0^1\rd\xi\int_0^\xi\rd\eta\ e^{\frac{\eta Q_s^2}{2}(\bx_1-\bx_{2})\cdot(\bx'_2-\bx'_1)}. \notag
\end{eqnarray}
\end{widetext}
The above result can be obtained by using the technique developed in many early works\cite{Gelis:2001da, Blaizot:2004wv, Dominguez:2008aa, Marquet:2010cf, Dominguez:2012ad, JalilianMarian:2004da}. The saturation momentum $Q_s^2$, which is proportional to $A^{1/3}$ with $A$ the number of nucleons, characterizes the density of target nuclei and it increases with the collisional energy. 

To reduce the number of integrations, we fix all the rapidities and set the rapidity of $\mathcal{Q}$ and $\mathcal{\bar Q}$ to be approximately equal. As usual, the $g\to \mathcal{Q}\mathcal{\bar Q}$ splitting function 
$\psi(\br)\psi^*(\br')\equiv\sum_{\lambda\alpha\beta}\psi^{\rT\lambda}_{\alpha\beta}(\br)\psi^{\rT\lambda*}_{\alpha\beta}(\br')=\frac{8\pi^2m_\mathcal{Q}^2}{p_g^+}\left[\frac{1}{2}K_1(m_\mathcal{Q}r)K_1(m_\mathcal{Q}r')\frac{\br\cdot\br'}{rr'}+K_0(m_\mathcal{Q}r)K_0(m_\mathcal{Q}r')\right]$
with $p_g^+$ the longitudinal momentum of the incoming gluon. To make further simplification, we set the longitudinal momentum fraction of $\mathcal{Q}$ and $\mathcal{\bar Q}$ with respect to the incoming gluon to be $\frac{1}{2}$ in the splitting function. 

In addition, we assume the momentum and coordinate distribution of the incoming gluon and quark inside the proton as the Gaussian type Wigner function $W({\bf b}, {\bf p}) =
\frac{1}{\pi^2} e^{-{\bb}^2/{B_p}-{\bp}^2/\Delta^2}$, where the parameters $B_p$ and $\Delta^2$ are the variances of the impact parameter $\bb$ and the transverse momentum $\bp$, respectively. This parameterization of incoming quark and gluon distributions can help us to perform some of the impact parameters and dipole size integrations analytically and allow us to carry out the rest of the integrals numerically. 

In the CEM, since the invariant mass of the heavy quark pair is integrated from the bare quark pair mass ($2m_{\mathcal{Q}}$) to the mass of the open heavy meson pair ($2M_{H}$), we should convolute the factor $\theta\left(\sqrt{M_H^2-m_{\mathcal{Q}}^2}-\Delta k_1\right)F_{\mathcal{Q}\mathcal{\bar Q}\rightarrow J/\psi}$ and integrate over the relative momentum $\Delta \bk_1$ to convert the produced $\mathcal{Q}\mathcal{\bar Q}$ into the corresponding heavy quarkonium with the probability $F_{\mathcal{Q}\mathcal{\bar Q}\rightarrow J/\psi}$. To further simplify the calculation, since the dominant contribution of the integration over $\Delta k_1$ comes from the region where $\Delta k_1 \sim m_\mathcal{Q}$\cite{Qiu:2013qka}, we assume that the threshold $\sqrt{M_H^2-m_{\mathcal{Q}}^2}$ is large enough ($1.4$ GeV for $J/\psi$) so we can integrate over $\Delta k_1$ up to infinity and obtain the delta function $(2\pi)^2\delta^{(2)}(\br-\br')$. We use such a crude approximation as a first step estimate. At last, the transverse momentum dependent production spectrum of the heavy quarkonium accompanied by a light quark in pA collisions reads
\begin{widetext}
\begin{eqnarray}
\frac{\rd N^{pA\rightarrow J/\psi qX}}{\rd^2\bk_1\rd^2\bk_2}&=&\prod_{i=1}^2W\left(\bb_i,\bp_i\right)\otimes\frac{\rd N^{gqA\rightarrow \mathcal{Q}\mathcal{\bar Q} qX}}{\rd^2\bk_1\rd^2\bk_2}F_{\mathcal{Q}\mathcal{\bar Q}\rightarrow J/\psi} \nonumber
\\
&=&\mathcal{N}\int\rd^2\br\prod_{i=1}^2\int\frac{\rd^2\bb_i\rd^2\br_i}{(2\pi)^{2}}\rd^2{\bp}_iW({\bb}_i,{\bp}_i)e^{-\ri(\bk_i-\bp_i)\cdot\br_i}\left<DDD|_{\br=\br'}\right>|\psi(r)|^2F_{\mathcal{Q}\mathcal{\bar Q}\rightarrow J/\psi}.
\end{eqnarray}
\end{widetext}

The $n$-th Fourier harmonics of the transverse momentum dependent differential spectrum is defined as\cite{Borghini:2001vi}
\begin{eqnarray}
\frac{\rd\kappa_{n}}{\rd k_1}\equiv k_1 \int d\phi_1\rd^2\bk_2e^{\ri n(\phi_1-\phi_2)}\frac{\rd N^{pA\rightarrow J/\psi qX}}{\rd^2\bk_1\rd^2\bk_2}, 
\end{eqnarray}
The $k_\perp$ dependent elliptic flow (the $2$nd Fourier harmonic) of the produced heavy quarkonium then can be obtained as follows\cite{CMS:2018xac}
\begin{eqnarray}
v_2(k_\perp)\equiv \frac{\frac{\rd\kappa_{2}}{\rd k_\perp}}{\frac{\rd\kappa_0}{\rd k_\perp} } \frac{1}{v_2[\text{ref}]},
\end{eqnarray}
where $v_2[\text{ref}]=\sqrt{\kappa_2[\text{ref}]/\kappa_0[\text{ref}]}$ is the transverse momentum integrated elliptic flow of the reference light quark which has been computed in Ref.~\cite{Davy:2018hsl}. Similarly, the integrated $v_2$ for heavy quarkonia can be written as $v_2\equiv (\kappa_2/\kappa_0)/v_2[\text{ref}]$.
 
It is interesting to notice that the four correlators inside $\left<DDD|_{\br=\br'}\right>$ cancel completely if we set the coordinate separation of the $\mathcal{Q}\mathcal{\bar Q}$ pair $r$ to $0$. Therefore, if we perform the small $r$ expansion, we can see the first nontrivial contribution comes at $r^2$ order and the mass dependence is associated with the $r$ integration. The heavy quark mass dependences cancel completely between $\kappa_2$ and $\kappa_0$ when we only compute $v_2$ up to the $r^2$ order.

As shown above, we have to evaluate a large number of integrations in order to numerically plot the elliptic flow of heavy quarkonia. Our strategy is to analytically integrate as many integrals as possible and numerically evaluate the remaining five dimensional integrations. 

\begin{figure}[tbp]
\begin{center}
\includegraphics[height=6.8cm]{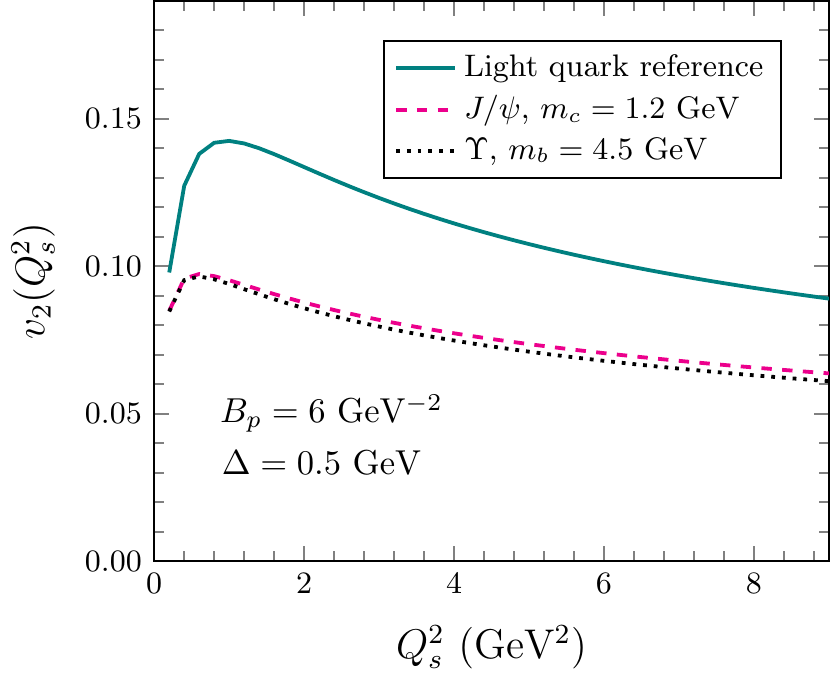}
\end{center}
\caption[*]{The integrated $v_2$ of $J/\psi$ and $\Upsilon$ compared with the $v_2$ of the reference light quark as function of the saturation momentum $Q_s^2$.}
\label{hq-qs}
\end{figure}

\begin{figure}[tbp]
\begin{center}
\includegraphics[height=6.9cm]{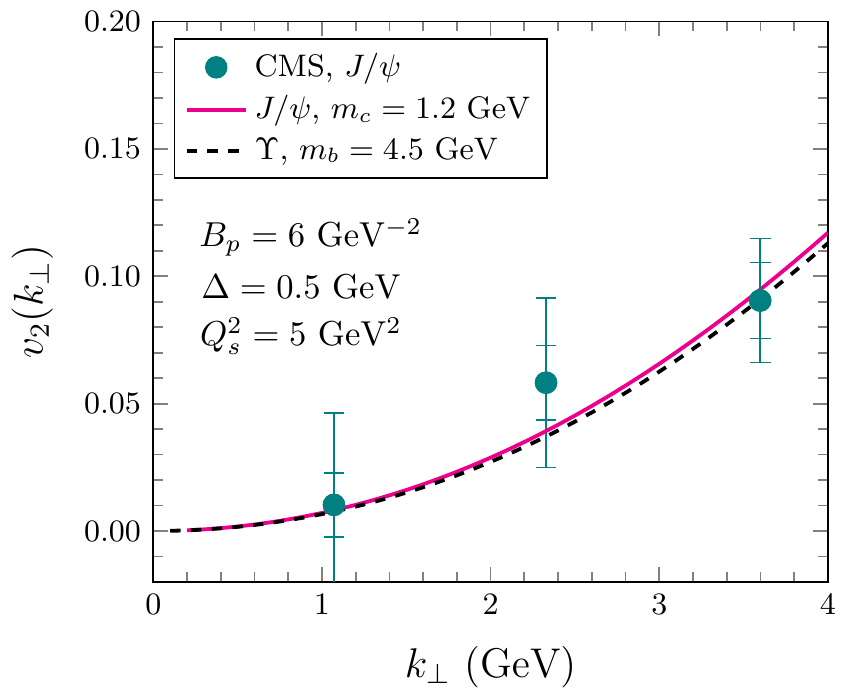}
\end{center}
\caption[*]{The $k_\perp$ dependent elliptic flow $v_2$ of $J/\psi$ as function of its transverse momentum $k_\perp$ compared with the CMS data\cite{CMS:2018xac} where both systematic (inner ones) and statistic (outer ones) error bars are shown. Our result is also consistent with the ALICE\cite{Acharya:2017tfn} data. In addition, as a prediction, the $v_2$ of $\Upsilon$ is also plotted in this figure.}
\label{hq-pt}
\end{figure}

\textit{3. Numerical results and comments} Using the aforementioned simplified model, we are able to compute the elliptic flow $v_2$ for heavy quarkonia, such as $J/\psi$ and $\Upsilon$ mesons. In Fig.~\ref{hq-qs}, we show that the integrated $v_2$ of $J/\psi$ and $\Upsilon$ comparing with the $v_2$ of light reference quark as functions of $Q_s^2$ in the CGC formalism. This plot shows that heavy quarkonia in the CGC formalism can typically have the $k_\perp$ integrated $v_2$ roughly between $5\% \to 10\%$, which is about $2/3$ of that for light reference quarks. Similar curves for light quarks can also be found in Refs.~\cite{Dusling:2017dqg, Dusling:2017aot}. It is important to note that, due to the splitting $g\to \mathcal{Q}\mathcal{\bar Q}$, the production mechanism of heavy quarkonia is generically different from that of light hadrons. We believe that this is the reason which leads to a slightly smaller $v_2$ for heavy quarkonia. In the meantime, the quarkonium mass dependence is rather weak for the elliptic flow if one compares between $J/\psi$ and $\Upsilon$. It will be very interesting to measure the $v_2$ of $\Upsilon$ in the near future. In Fig.~\ref{hq-pt}, excellent agreement is found by comparing our calculation of $v_2 (k_\perp)$ for $J/\psi$ to the CMS data with the parameter consistent with Ref.~\cite{Davy:2018hsl}. Here we use a slightly larger $Q_s^2=5 \text{GeV}^2$ for the LHC instead of $Q_s^2=4 \text{GeV}^2$ for RHIC. 

\textit{4. Conclusion and outlook} As a conclusion, let us make some further comments on the consequence of this work. 
\begin{itemize}
\item First of all, as we have demonstrated above, the heavy quarkonia can have significant elliptic flow due to color interactions and transitions which have little mass dependence. Intuitively, this can be understood as the cancellation of mass dependence between the anisotropic spectrum $\kappa_2$ and the isotropic spectrum $\kappa_0$, thus $v_2$ contains little mass dependence. This allows us to predict that the $\Upsilon$ meson should also have similar size of elliptic flow at RHIC and the LHC, although it is much heavier than the $J/\psi$ meson. This prediction could be tested in future measurements. 
\item Furthermore, instead of integrating over the relative transverse momentum of heavy quark pairs, we can integrate over the momentum of $\mathcal{\bar Q}$ and measure the outgoing $\mathcal{Q}$. This allows us to generalize the above calculation and compute the elliptic flow for open charm particles, namely the $D^0$ meson. The numerical evaluation may be more demanding, but we expect the corresponding $v_2$ for $D$ mesons should lie in the similar range as that of $J/\psi$. 
\item In addition, instead of using the CEM, one could also compute the elliptic flow for heavy quarkonia in more sophisticated models by separating the color singlet states from the color octet states. Nevertheless, since the elliptic flow is computed from the ratio of $\kappa_2$ and $\kappa_0$ where most of the detailed information of the hadronization from heavy quark pairs to physical quarkonia cancels, we expect that our prediction for $v_2$ should be robust.  
\item Last but not least, the framework employed in this paper is consistent with previous calculations on the spectra of $J/\psi$ and $\Upsilon$ mesons in both $pp$ and $pPb$ collisions\cite{Watanabe:2015yca}. It is worth noting that one can describe both the elliptic flow $v_2$ and the nuclear modification factor $R_{pPb}$ for heavy quarkonia in the low transverse momentum region within this framework. A similar but more comprehensive description of $J/\psi$ production in $pp$ and $pPb$ collisions can be also found in Refs.~\cite{Ma:2014mri, Ma:2015sia}.
\end{itemize}

In this paper, we have computed the elliptic flow for heavy quarkonia and found 
excellent agreement with the $J/\psi$ data measured at the LHC. This suggests that the observed large $v_2$ for $J/\psi$ 
at the LHC can be naturally explained as the initial state effect in the CGC formalism. 

Due to the complexity of this problem, a number of approximations have been made in order to simplify the calculation before we can perform the numerical calculation. Nevertheless, we expect the main feature of our results will retain in a more complete calculation. We leave such study and the detailed derivation as well as the calculation of the $v_2$ of $D^0$ mesons to a future work.

\textit{Acknowledgments:} We thank Z. W. Lin,  A. Mueller, F.Q. Wang and F. Yuan for useful discussions and comments. This material is based on the work supported by the Natural Science Foundation of China (NSFC) under Grant Nos.~11575070, 11775095, 11890711 and 11375072. CM and SYW are supported by the Agence Nationale de la Recherche under the project ANR-16-CE31-0019-02.

\end{document}